\documentclass[twocolumn,  twocolappendix]{aastex631}

\shorttitle{New Pulsar Discovered in Glimpse-C01}
\shortauthors{McCarver et al.}

\newcommand{\glimpse}{GLIMPSE-C01}

\begin{document}

\title{A VLITE Search for Millisecond Pulsars in Globular Clusters: Discovery of a Pulsar in \glimpse}

\author[0009-0008-0527-4082]{Amaris V. McCarver}
\affiliation{Department of Physics \& Astronomy, Texas Tech University, Box 41051, Lubbock TX, 79409}
\affiliation{Naval Research Enterprise Internship Program (NREIP), U.S. Naval Research Laboratory, Washington, DC, 20375,  USA}

\author{Thomas J. Maccarone}
\affiliation{Department of Physics \& Astronomy, Texas Tech University, Box 41051, Lubbock TX, 79409}

\author[0000-0001-5799-9714]{Scott M.~Ransom}
\affiliation{National Radio Astronomy Observatory, 520 Edgemont Rd., Charlottesville, VA, 22903, USA}

\author[0000-0001-6812-7938]{Tracy E.\ Clarke}
\affiliation{U.S.\ Naval Research Laboratory,  4555 Overlook Ave SW,  Washington,  DC 20375,  USA}

\author[0000-0002-1634-9886]{Simona Giacintucci}
\affiliation{U.S.\ Naval Research Laboratory,  4555 Overlook Ave SW,  Washington,  DC 20375,  USA}

\author[0000-0002-5187-7107]{Wendy M. Peters}
\affiliation{U.S.\ Naval Research Laboratory,  4555 Overlook Ave SW,  Washington,  DC 20375,  USA}

\author[0000-0003-3272-9237]{Emil Polisensky}
\affiliation{U.S.\ Naval Research Laboratory,  4555 Overlook Ave SW,  Washington,  DC 20375,  USA}

\author[0000-0003-1991-370X]{Kristina Nyland}
\affiliation{U.S.\ Naval Research Laboratory,  4555 Overlook Ave SW,  Washington,  DC 20375,  USA}

\author{Tasha Gautam}
\affiliation{Max-Planck-Institut für Radioastronomie,  Auf dem H\"ugel 69,  D-53121,  Bonn,  Germany}
\affiliation{National Radio Astronomy Observatory, 520 Edgemont Rd., Charlottesville, VA, 22903, USA}

\author{Paulo C. C. Freire}
\affiliation{Max-Planck-Institut für Radioastronomie,  Auf dem H\"ugel 69,  D-53121,  Bonn,  Germany}

\author{Blagoy Rangelov}
\affiliation{Department of Physics, Texas State University, 601 University Drive, San Marcos, TX 78666, USA}

\begin{abstract}
We present results from a search for pulsars in globular clusters, including the discovery of a new millisecond pulsar in the stellar cluster \glimpse. We searched for low frequency radio sources within 97 globular clusters using images from the VLA Low-band Ionosphere and Transient Experiment (VLITE) and epochs 1 and 2 of the VLITE Commensal Sky Survey (VCSS). We discovered 10 sources in our search area, four more than expected from extragalactic source counts at our sensitivity limits. The strongest pulsar candidate was a point source found in \glimpse\ with a spectral index $\sim -2.6$, and we present additional measurements at 0.675 and 1.25 GHz from the GMRT and 1.52 GHz from the VLA which confirm the spectral index. Using archival Green Bank Telescope S-band data from 2005, we detect a binary pulsar with a spin period of 19.78~ms within the cluster. Although we cannot confirm that this pulsar is at the same position as the steep spectrum source using the existing data, the pulse flux is consistent with the predicted flux density from other frequencies, making it a probable match. The source also shows strong X-ray emission, indicative of a higher magnetic field than most millisecond pulsars, suggesting that its recycling was interrupted.  We demonstrate that low frequency searches for steep spectrum sources are an effective way to identify pulsar candidates, particularly on sightlines with high dispersion.

\end{abstract}

\keywords{Millisecond pulsars (1062),  Globular star clusters (656),  Galactic radio sources (571)}

\section{Introduction} \label{sec:intro}

Pulsars are rotating neutron stars (NSs) that emit electromagnetic radiation from their poles,  with radio waves being the most common. This radio emission is broadband, and it typically has a steep spectral index, which means that the radio emission is stronger at lower frequencies (\citealt{Bates2013}). The pulse period ($P$) reflects the pulsar's rotation; its time derivative ($\dot{P}$), always positive, its slowdown. Using $P$ and $\dot{P}$ we can estimate the pulsar's characteristic age and its surface magnetic field.

Millisecond pulsars (MSPs) are pulsars with extremely fast and stable rotations, i.e., with very small values 
of $P$ and $\dot{P}$ that indicate very large characteristic ages (of the order of $10^9 \, \rm yr$) and comparably small values of magnetic field ($\sim 10^8 \, \rm G$, four orders of magnitude smaller than for
"normal" pulsars). They are created through spin-up processes which require accretion of mass from a companion star \citep{2023pbse.book.....T}; during this accretion stage the system is a low-mass X-ray binary (LMXB).

Given their large stellar densities, globular clusters are predicted to be an excellent environment for the formation of MSPs. The stellar densities are so large that many NSs, which in the Galaxy would never be recycled, can acquire a companion via binary exchange encounters. If these companions fill their Roche lobe they start transferring matter to the NS. In globular clusters such LMXBs are $\sim 10^3$ times more numerous per unit of stellar mass than in the Galaxy \citep{1975ApJ...199L.143C}. Given that MSPs are produced via spin-up in LMXBs \citep{1976ApJ...207..574S,1982Natur.300..728A}, this should result in a large number of MSPs, which has been confirmed since the discovery of PSR~B1821$-$24A in the globular cluster M28 \citep{1985AJ.....90..606H,1987Natur.328..399L}, and more recently, by clearly established statistical excesses of millisecond pulsars in globular clusters \citep{2008IAUS..246..291R,2013IAUS..291..243F}\footnote{At the time of writing, a total of 305 pulsars have been found in 40 globular clusters, see Freire's catalog at \url{https://www3.mpifr-bonn.mpg.de/staff/pfreire/GCpsr.html}}.

The typical method for finding pulsars is to use time domain surveys, where radio spectra (typically with about a thousand channels) are sampled and recorded at a few tens of kHz, which results in very large data volumes. The regular pulsations of radio pulsars are then searched in these data typically using Fourier transform methods \citep{2002AJ....124.1788R,2018ApJ...863L..13A}. These can be very computationally intensive, especially if the aim is to detect fast-spinning pulsars in compact or highly eccentric orbits, for which the surveys are clearly incomplete. The problem is compounded, in the case of radio interferometers, by the fact that hundreds of such time-domain beams must be formed in order to cover even a relatively small region of interest like a globular cluster, resulting in data volumes that simply cannot be stored \citep{2021MNRAS.504.1407R}.
In addition, dispersive smearing and especially multi-path propagation ("scattering") can render pulsars undetectable, especially the faster ones. This problem affects mostly lower radio frequencies, where pulsar radio emission is strongest.

Imaging surveys at low frequencies for steep spectrum radio sources offer an effective alternative to time-domain searches because although they may have any spectral index, pulsars are one of the only objects expected to have extremely steep spectral indices \citep{Jankowski2017}. Imaging surveys are impervious to the issues that hinder time-domain surveys, like extreme accelerations and scattering; this means that they retain sensitivity to potentially very exciting systems that remain beyond the reach of time domain surveys. The downside is that, apart from the position of the source and its flux density, they offer no characterization of the new pulsar, which still requires a time-domain survey. Nevertheless, by identifying the source and determining its position, they can greatly focus the survey (especially in the case of interferometers like MeerKAT and the future SKA) and help prioritize computing resources. 

As an example, the first MSP (PSR~B1937+21) was found after targeting of a steep spectrum, polarized radio source \citep{1982Natur.300..615B}. A more relevant example for this work was the first pulsar discovered in a globular cluster: this was found in a Very Large Array (VLA) imaging survey of globular clusters by \cite{1985AJ.....90..606H}, which was motivated by the early LMXB discoveries in globular clusters. It was by following up on the brightest of these sources with a focused time-domain survey that \cite{1987Natur.328..399L} confirmed a 3.05-ms pulsar, PSR~B1821$-$24A. Later imaging surveys like that of \cite{2000ApJ...536..865F,2017MNRAS.468.2526B} have revealed steep-spectrum radio emission in the cores of Terzan 5, NGC~6544 and Liller 1, and in the Galactic Center region, well before those pulsars were found (and in the case of Liller 1, no pulsations have been found yet, possibly because of strong scattering).

Recently \cite{Gautam2022} present a search for continuum and pulsed emission in four globular clusters at 400 MHz and another four at 650 MHz using the upgraded Giant Metrewave Radio Telescope (uGMRT) in India.  They specifically targeted clusters with previously known pulsars.  The results of their study were the identification of a new MSP, and 8 additional continuum sources not coincident with any of the known pulsars in the clusters.  Many of these sources were found on the fringes of the clusters, outside the area typically searched by time domain searches. Three of the pulsars in the clusters had broad profiles, and the continuum emission was much stronger than the pulsed flux, suggesting that in these pulsars the radio emission is mostly unpulsed. They suggest that low-frequency continuum searches thus offer a powerful tool to increase the census of known pulsars in globular clusters.

In this paper we present a large search for pulsar candidates in globular clusters using low frequency radio continuum images.
Throughout this paper, the radio spectral index $\alpha$ is defined according to $S_{\nu} \propto \nu^\alpha$, where $S_\nu$ is the measured flux density at the observing frequency $\nu$. 

\section{Globular Cluster Sample}
\label{sample}

Radio interferometers operating at low frequencies ($\nu < 1$ GHz) typically have relatively crude resolutions ($\sim 5^{\prime\prime} - 30^{\prime\prime}$ or worse) relative to instruments more typically used to find new pulsars at higher frequency.  As a result, for clusters already well-studied by pulsar timing experiments, it could be difficult to distinguish known pulsars from new candidate pulsars based on low-frequency continuum images. To avoid this difficulty, we compiled a sample of 121 globular clusters that lacked previously identified pulsars using the Harris and Freire catalogs (\citealt{Harris1996}, 2010 edition). Of these, 97 are at declinations higher than $-40^\circ$, and thus visible from the instruments in the northern hemisphere.  Detections of radio point sources without known identifications, particularly those that are very steep spectrum, can thus be identified as pulsar candidates.

\section{Data}
\subsection{VLITE}\label{sec:data}

Our initial search of the 97 globular clusters started with data from the VLA Low-band Ionosphere and Transient Experiment (VLITE)\footnote{\url{https://vlite.nrao.edu/}}. VLITE is a commensal system on the National Radio Astronomy Observatory's Karl G. Jansky Very Large Array (NRAO VLA) which records data at 340 MHz during nearly all regular VLA observations \citep{clarke2016,polisensky2016}. Operating on 16-18 antennas since August of 2017,  VLITE records approximately 6000 hours of data per year, and its archive covers $98\%$ of the sky north of declination $-40^\circ$ to a depth of $33$ minutes.  

All VLITE data are processed within a few days of observation by a dedicated calibration and imaging pipeline which combines python with standard tasks found in AIPS \citep{aips} and Obit \citep{obit}. Complex gains and bandpass solutions are found using one or more observations of 6 standard calibrators: \texttt{3c48, 3c138, 3c147, 3c286, 3c295}, and \texttt{3c380}. The calibrator must be observed within 24 hours of the target data. The pipeline uses full-field calibrator models that have been scaled to match the flux scale models of \citet{pb2017} at the central VLITE frequency. After primary calibration, the data are flagged to remove interference,  and phase calibrated using a global sky model based on the NVSS \citep{Condon1998}. Finally, each target dataset undergoes several rounds of imaging and self-calibration to produce a final, single-frequency image. Further details of the calibration and imaging process are given in \citet{polisensky2016}. All VLITE images are archived and the sources are cataloged into a dedicated SQL source database \citep{vdp2019}.  

A special on-the-fly (OTF) correlation mode was enabled to allow observations during the VLA Sky Survey \citep[VLASS]{lacy+20} creating the complementary VLITE Commensal Sky Survey (VCSS)\footnote{Epoch 1 of the VCSS Bright Source Catalog is available at \url{https://cirada.ca/vcsscatalogue}}.  These data are processed using a modified version of the standard VLITE calibration pipeline as described in \citet{Peters+2023}.

For this study, we extracted 0.25$^{o}$ square image cutouts from the first and second epochs of VCSS (referred to herein as VCSS1 and VCSS2, respectively) at the positions of the 97 clusters with a resolution of $\sim15^{\prime\prime} - 25^{\prime\prime}$,  and a typical $1\sigma$ root-mean-squared (RMS) noise of $\sim3-5$ mJy bm$^{-1}$.

We also searched the non-OTF VLITE archive (referred to herein as VLITE images) for existing images from observations between July 2017 and July 2022 which covered the cluster positions.  Where multiple images at a variety of different pointings were available, they were corrected for the standard VLITE asymmetric primary beam response and combined in the image plane using standard tasks in the Obit MosaicUtil python package to generate a deeper image at the cluster position. For 8 of the clusters, multiple datasets at matched pointings were available. The pipeline calibrated data for these were combined and re-imaged using either the Obit imaging task \texttt{MFImage} or \texttt{WSCLEAN} \citep{wsclean} to generate a more sensitive final image. Combining these two methods, we produced $0.25^{o}$ square images for 33 clusters at resolutions of $\sim5 - 25^{\prime\prime}$, and RMS noise values $\sim5$ mJy bm$^{-1}$ or better. 

\begin{figure*}
\centering
\includegraphics[clip=true, trim=0cm 0cm 0.25cm 0cm, width=0.75\textwidth]{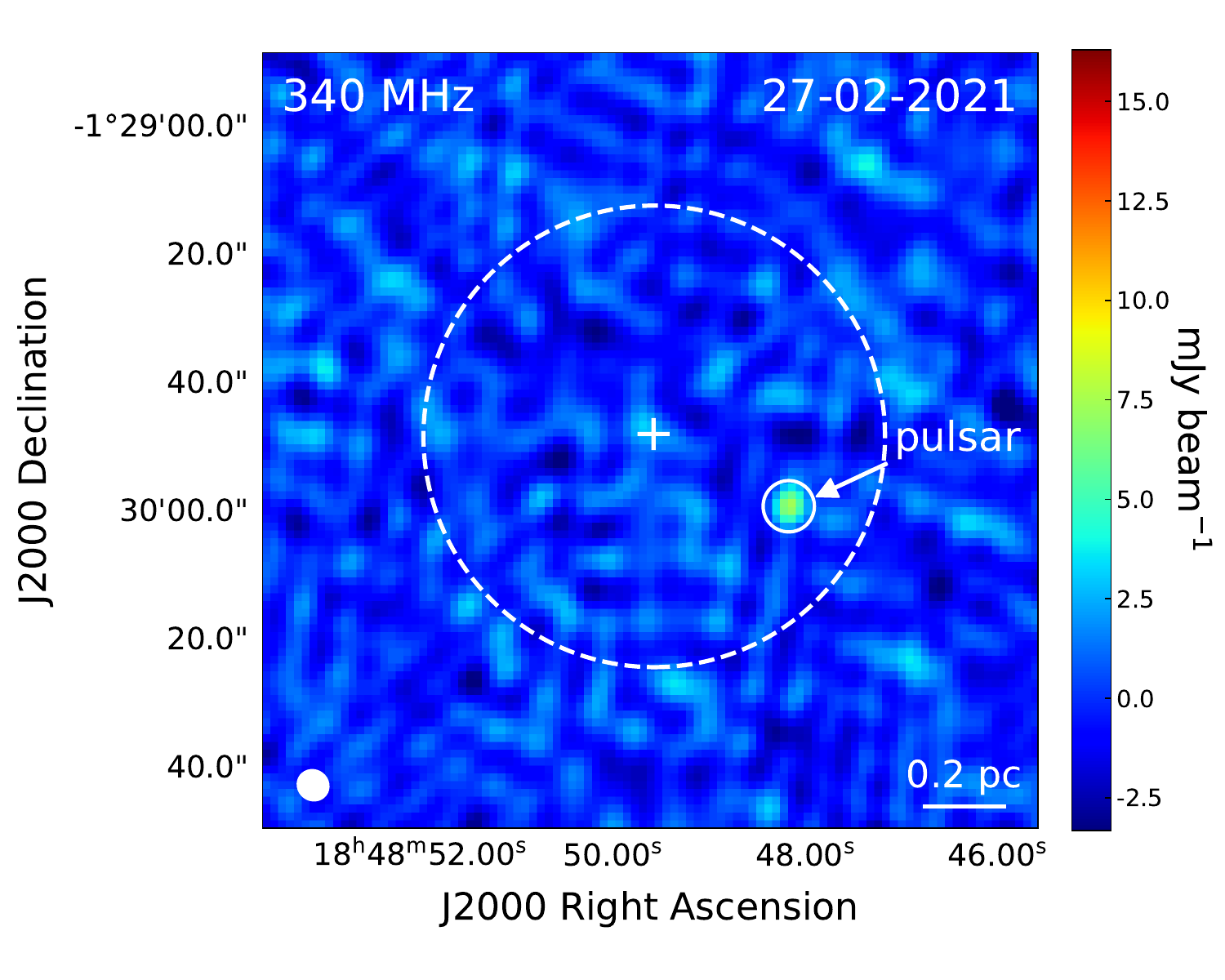}
\caption{VLITE 340 MHz image of \glimpse\ from 27 February 2021. The clean beam is shown as a white ellipse in the lower left corner and has dimensions of  $5.0^{\prime \prime} \times 4.7^{\prime \prime}$ with a position angle of $52^{\circ}$.  The cross denotes the central position of \glimpse.  The dashed white circle shows the core radius of 36$^{\prime \prime}$.  The location of the pulsar candidate is shown with a solid white circle.  A scalebar indicating a linear size of 0.2~pc (12.5$^{\prime \prime}$), assuming a distance to \glimpse\  of 3.3~kpc, is shown in the lower right corner.
}
\label{fig:VLITE_Glimpse}
\end{figure*}

We used PyBDSF \cite{pybdsf} to catalog all VLITE/VCSS sources at $>3\sigma$ inside the cluster radius (the larger of either half-light or core radius was chosen for each cluster, as indicated in Table~\ref{tab:sample}). In total, 10 sources were identified which were either at $>5\sigma$ significance or between $3-5\sigma$ and matched a known source at another radio frequency. A list of the sources, their fluxes, spectral indices, and any known identifications are given in Table~\ref{tab:Sources1}. The source in \glimpse\ (RA 18:48:48.1, Dec -01:29:58) was the steepest spectrum source and thus the strongest pulsar candidate. We show the VLITE image of \glimpse\ with the pulsar labeled in Figure~\ref{fig:VLITE_Glimpse} and discuss this source in detail in Section~\ref{sec:glimpse}.

Detections of sources in VLITE data were compared to available sky surveys with sub-arcmin resolutions at frequencies between 150 MHz and 3 GHz to derive a spectral index for each detection, assuming a simple power law fit.  These included the Very Large Array Sky Survey (VLASS, 3GHz; \citealt{lacy+20}), the NRAO VLA Sky Survey (NVSS, 1.4GHz; \citealt{Condon1998}), the Faint Images of the Radio Sky at Twenty-cm (FIRST, 1.4 GHz; \citealt{1995ApJ...450..559B}), the Rapid ASKAP Continuum Survey (RACS-low, 887.5 MHz; \citealt{Hale_2021}), the Sydney University Molonglo Sky Survey (SUMSS, 850 MHz; \citealt{Mauch2003}), 
\citealt{Hurley-Walker}), 
and the TIFR GMRT Sky Survey-redux (TGSSr, 150 MHz; \citealt{refId0}).

Our strongest candidate, \glimpse, was followed up with both archival and new radio observations as well as archival X-ray observations as discussed below.

\subsection{Additional Data for \glimpse\ }
\label{sect:additional}

For the cluster \glimpse\ 
we analyzed archival and new observations at a variety of additional wavelengths to followup on the candidate pulsar.  We summarize these data here. We note that the RACS-low image in the regions of \glimpse\ is too distorted by nearby Galactic structures to allow us to use it to set a meaningful flux constraint. Table~\ref{tab:GlimpseFlux} presents all fitted flux density measurements from archival and new observations for the \glimpse\ pulsar candidate.

\subsubsection{JVLA}

We observed \glimpse\ with the JVLA under project 23A-277 (PI: Peters, W.) using the L-band receiver for 1 hour on each of 07 May 2023 in B-configuration, and 14 August 2023 and 18 September 2023 in A-configuration. The observations were made using a standard frequency setup with 16 x 64 MHz IFs, at a central frequency of 1520 MHz. The data were calibrated using the JVLA CASA pipeline, with \texttt{3c286} as the primary calibrator for gains, bandpass, and delays, and observations of the source J1851+0035 were used to calibrate complex gains.  The A-configuration data from August and September were combined after the primary calibration was complete. In order to reduce the effect of extended Galactic emission, baselines shorter than 5 k$\lambda$ in the B-configuration data, and 14 k$\lambda$ in the A-configuration were removed.  Each of the datasets were then imaged and self-calibrated in phase using the task \texttt{MFImage} in Obit.  Problematic field sources with sidelobes extending through the pulsar position were identified and peeled using the software algorithm described in ~\cite{Cotton2021}.  

The final image from the May data has a resolution of $4.6^{\prime\prime} \times 3.3^{\prime\prime}$ at a position angle of $59.7^{o}$ at a central frequency of 1520 MHz. The local noise is 54 $\mu$Jy bm$^{-1}$, and a source at the pulsar position is detected at $4.8\sigma$ significance, with a flux density of S$_{1520} = 269 \pm 54\, \mu$Jy.  The A-configuration data were split into two frequency sub-bands and a final image made for each.  The lower frequency A-configuration image has a resolution of $1.33^{\prime\prime} \times 1.0^{\prime\prime}$ at a position angle of $65.3^{o}$ and a local rms of $40 \mu$Jy bm$^{-1}$ at a central frequency of 1296 MHz.  A source at the pulsar position is detected at $6.2\sigma$ significance with a flux density of $S_{1296} = 253 \pm 70\, \mu$Jy.   The higher frequency A-configuration image has a resolution of $0.9^{\prime\prime} \times 0.7^{\prime\prime}$ at a position angle of $84.7^{o}$ and a central frequency of $1840$ MHz. The local noise RMS is $29\, \mu$Jy bm$^{-1}$, and the pulsar is just detected at a significance of $3.0\sigma$, and a flux density of $89 \pm 51\, \mu$Jy.

\subsubsection{Upgraded GMRT} 

We observed \glimpse\ with the uGMRT in Band 4 (550--950 MHz) on 2022 July 10 and on 2023 January 2, for 1 hour on each day, including calibration overheads 
(projects DDTC234 and 43$_{-}$049). 
We observed the cluster in Band 3 (250--500 MHz) and Band 5 (950--1460 MHz) on 2023 January 2 for 1 hour in each band (Project 43$_{-}$049).
All data were collected in spectral-line mode using the wide-band back-end with a bandwidth of 400 MHz, 8192 frequency channels, and an integration time of 2.6 seconds. At all frequencies, the source \texttt{3c286} was observed as bandpass and absolute flux density calibrator. 

We reduced the data using the Source Peeling and Atmospheric Modeling \citep[\texttt{SPAM};][]{2009A&A...501.1185I} pipeline\footnote{http://www.intema.nl/doku.php?id=huibintemaspampipeline}. 
For each observation, the full-band dataset was first divided into 6 narrower sub-bands. Each sub-band was then processed independently by the \texttt{SPAM} pipeline, adopting a standard calibration scheme that consists of bandpass and complex gain calibration, as well as flagging of radio frequency interference. The flux density scale was set using
\cite{2012MNRAS.423L..30S}. 
After initial calibration, direction-independent self-calibration was applied to the target data, followed by direction-dependent self-calibration. We used the VLITE 340 MHz image as initial sky model for the phase self-calibration process. For each observations, we imaged the final self-calibrated sub-bands together using joint-channel deconvolution in \texttt{WSClean} \citep{wsclean} and produced final images at the central frequencies of 400 MHz, 675 MHz and 1280 MHz. In the imaging process, we used uniform weights and excluded baselines shorter than 5 k$\lambda$ to minimize the effect of diffuse Galactic emission in the field. The final images have a resolution of $9^{\prime\prime}\times5^{\prime\prime}$ (Band 3), $4^{\prime\prime}\times4^{\prime\prime}$ (Band 4) and $3^{\prime\prime}\times2^{\prime\prime}$ (Band 5) and an rms of 230 $\mu$Jy bm$^{-1}$, 55 $\mu$Jy bm$^{-1}$, and 78 $\mu$Jy bm$^{-1}$, respectively. The systematic amplitude uncertainty was assumed to be 15$\%$ at all frequencies.

\subsection{GBT}

\glimpse\ was observed for 7.36\,hours with the Green Bank Telescope on 2005 August 13 (Project GBT05B$-$045, PI: B.~Jacoby) using the Spigot pulsar instrument \citep{spigot05} at a central observing frequency of 1850\,MHz and the S-band receiver. As the Spigot was an autocorrelation spectrometer, 1024 lags were saved every 81.92\,$\mu$s, which were then converted into spectra with 1024 channels using an early version of {\sc PRESTO} \citep{presto}. Since the lower $\sim$270\,MHz of the sampled band was below the filtered cutoff of the S-band receiver,  the lowest 352 channels were discarded, and the remaining 672 channels were saved as SIGPROC-style filterbank data with 8-bits per sample, giving 525\,MHz of usable bandwidth centered at 1987.5\,MHz.

\begin{figure*}
\centering
\includegraphics[clip=true, trim=0cm 0cm 0cm 0cm, width=0.93\textwidth]{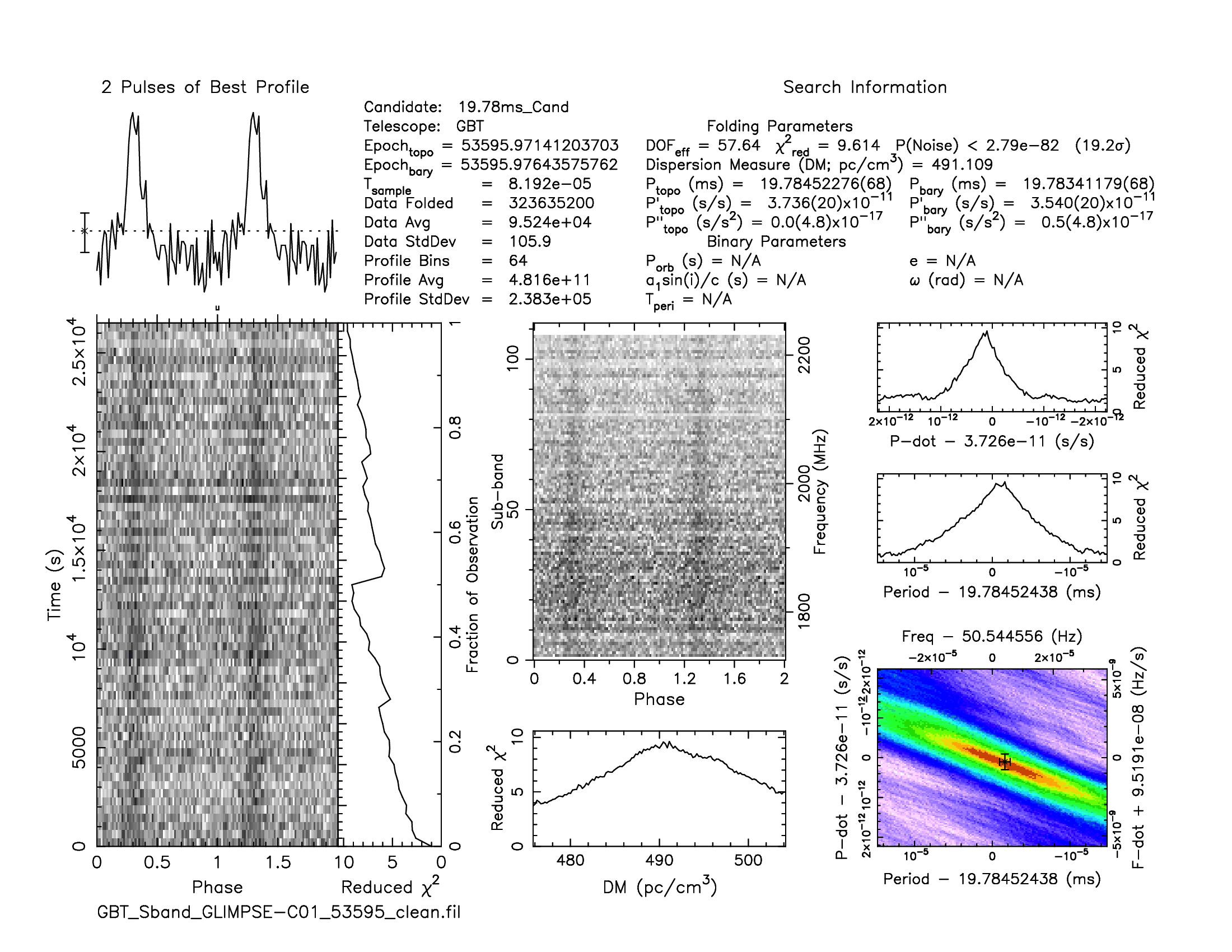}
\caption{Detection plot of the new pulsar in \glimpse\ as seen in archival GBT data from 2005. This is a standard pulsar candidate plot from the PRESTO routine {\tt prepfold}. Grey-scale portions of the image show the intensity of the pulsar emission after folding at the detected spin period as functions of pulse phase and (left) time of the observation and (center) observing frequency. We compute significance statistics of the detection by calculating the reduced-$\chi^2$ statistic for a model assuming no pulsations. The integrated pulse profile is showed at the top left.}
\label{fig:GBT_pulsar}
\end{figure*}

These data were searched for pulsars by the original observers, as well as independently by one of us (SMR), in 2005$-$2006 with no pulsations detected, and were then archived onto an external hard drive and stored on a dusty office shelf for $\sim$17\,years. After the radio detection of a point source within the cluster, we re-searched the data over a dispersion measure (DM) range from 0$-$1000\,pc\,cm$^{-3}$, using full acceleration searches that allowed {\sc PRESTO}'s {\tt accelsearch} to detect signals that drifted by up to 200 Fourier frequency bins during the observation \citep[i.e.~$zmax = 200$;][]{rem02}. We detected a strong, binary pulsar signal with spin period of 19.784\,ms and acceleration of -0.53\,m\,s$^{-2}$ at a DM of $\sim$490\,pc\,cm$^{-3}$ (see Fig.~\ref{fig:GBT_pulsar}). As this is the first pulsar detection within the cluster \glimpse\ we designate this pulsar as PSR\ J1848-0129A. We note that the pulsar signal was likely missed in the earlier searches of the GBT data since full acceleration searches were not typically conducted over wide DM ranges for long observations due to computational complexity. With today's computing resources, such searches are much easier.

\subsection{Chandra}
We used archival data of \glimpse\ taken with with {\it Chandra} (PI Rangelov). The program was split into six observations (ObsIDs 21641, 21642, 21643, 21644, 21645, 21646), each 30\,ks, taken over a period spanning 17 months from 2019 June 23 to 2020 November 19. The data were taken with the ACIS-I instrument operated in ``VeryFaint'' Timed exposure mode. We processed the data using the {\it Chandra} Interactive Analysis of Observations (CIAO\footnote{\url{http://cxc.cfa.harvard.edu/ciao/}}) software version 4.14 and {\it Chandra} Calibration Database version 4.10.2. The data have been restricted to the energy range between 0.5 and 7\,keV and filtered in three energy bands, 0.5--1.2\,keV (soft),  1.2--2\,keV (medium), and 2--7\,keV (hard). We used the CIAO's Mexican-hat wavelet source detection routine \texttt{wavdetect} \citep{Freeman2002} to create source lists. In order to find fainter point sources,  all six datasets were merged\footnote{Standard CIAO procedures found at \url{http://cxc.harvard.edu/ciao/threads/wavdetect_merged/} were followed to merge the data.} prior to running \texttt{wavdetect}. We aligned all {\it Chandra} datasets with respect to the first observation (ObsID 21641) using the CIAO scripts \texttt{wcs\_match} and \texttt{wcs\_update}. We used an exposure-time-weighted average PSF map in the calculation of the merged PSF. Taking into account the new aspect solutions, the observation event files were merged into one event file using \texttt{merge\_obs}. We detected a total of 21 cluster X-ray sources in the merged data. The \texttt{srcflux} CIAO tool was then run individually on each observation (using the coordinates found by \texttt{wavdetect}).

The X-ray spectra were fit using XSPEC (Arnaud 1996). We used an absorbed power law (PL) with The Tuebingen-Boulder ISM absorption model (\texttt{TBabs}), with the abundances set to \texttt{wilm} \citep{ Wilms2000}. The statistic used to analyze the fit was c-stat. The data were restricted to $0.5-8$\,keV, and the hydrogen column density was frozen to $n{\rm H} = 4\times10^{22}$\,cm$^{-2}$, which is the average for the cluster (see \citealt{Hare2018}). The fit resulted in $\Gamma = 0.9\pm0.8$ and flux of $F_X = (4.9\pm1.5)\times10^{-15}$\,erg/s/cm$^2$. The c-Statistic was 34.87 using 30 bins, with a null hypothesis probability of $7.73\times10^{-1}$ with 27 degrees of freedom.

\citet{Hare2018} performed the analysis of the {\it HST} images of \glimpse\ taken with WFC3/UVIS and IR, and the astrometric alignment between the {\it HST} and {\it Chandra} data. Here we use the same astrometric results. For both the {\it HST} and {\it Chandra} data, stars were matched from the Two Micron All-Sky Survey catalog \citep[2MASS;][]{2mass} to stars in the field of view of the observatories (for more details, see \citealt{Hare2018}). The astrometric correction was estimated to be 0.1”. We have identified two NIR counterparts in the {\it HST} data for the X-ray source at the radio position of the pulsar candidate. The magnitudes and offsets are listed in Table~\ref{tab:counterparts}. 


\section{Results\label{sec:results}}

In total we identify 10 sources in the VLITE and/or VCSS data for our sample of 97 clusters. Half of the sources were detected by VLITE at $>5\sigma$ significance; the remainder were detected between $3-5\sigma$ at the position of known sources in other radio catalogues. We briefly discuss each of the sources below and summarize their properties in Table~\ref{tab:Sources1}. 

We performed a weighted fit using archival and new flux density measurements to determine the spectral index for each source. We assigned weights to each measurement equal to the inverse of the fractional flux density uncertainty, giving greater importance to more precise measurements. In cases where the source was visible in the archival image but not in the corresponding survey catalog, we obtained the image data and performed our own measurement using PyBDSF at the VLITE location of the source. 
 
The resulting flux densities and fitted power-law spectra are presented in Figure ~\ref{fig:spectra}. Archival data is plotted in black, while new data presented in this work is shown in blue. This approach allowed us to obtain consistent and reliable spectral indices for all sources in our sample.

\subsection{Terzan 4}

Terzan 4 lies at a distance of  7.2 kpc from the Sun, and the core radius is $0.90^\prime$. The absolute visual magnitude is -4.48 magnitudes (\citealt{Harris1996}, 2010 edition).

We identified a single bright radio source within the half-light radius of Terzan 4, which we observe in both the first and second epochs of the VCSS. This source matches cataloged observations in TGSSr, RACS-low, NVSS, and the first and second epochs of VLASS. Although it is not included in the SUMSS catalog, there is a clearly visible source at this position. We have used PyBDSF to fit the source, measuring a flux density of 147 ± 7 mJy and a peak signal-to-noise ratio of 38. The source spectrum is well-described by a simple power-law with a spectral index of $\alpha \sim -0.9$, as shown in Figure~\ref{fig:spectra}.

\subsection{NGC 5466}
NGC 5466 is at a distance of 16.0 kpc from the Sun, and the core radius is $1.43^\prime$.  The absolute visual magnitude being -6.98 magnitudes (\citealt{Harris1996}, 2010 edition).

No source was detected in either VCSS1 or VCSS2. The cluster lies at a separation of $\sim21^\prime$ from the source J1407+2827 which is frequently observed as a calibrator at higher frequencies by the VLA, and for which VLITE has hundreds of archival images.  We convolved 128 of the highest quality images taken in A-configurations between 2018 and 2022 to a common resolution of 6.8" and combined them in the image plane.  After primary beam correction, the final rms at the cluster position is 0.23 mJy bm$^{-1}$.  We detect a point source with a total flux density of $1.8 \pm 0.3$ at a position within the core radius of NGC 5466.   

Although we found no matches for this source in any other published radio catalogs, a faint source is visible in the RACS-low image at the same position.  Using PyBDSF we are able to fit the source as a single component with a flux density of $1.6 \pm 0.6$ mJy and a peak SNR of 4.6.  It is similarly bright at 150 MHz ($2.1\pm0.2$ mJy) in the LOFAR Two-Meter Sky Survey (LoTSS).  These measurements are very similar to the VLITE flux density, suggesting that this is a nearly flat spectrum source in the 150-900 MHz range.  It is absent in the VLASS catalog \citep{Gordon_2021} (although visual inspection shows a source at about 0.5 mJy in all three epochs), indicating a steeper spectrum above 900 MHz.  The source could either be a Gigahertz-peaked pulsar, or, more likely, a background active galactic nucleus.

\subsection{Pal 11}
Pal 11 is at a distance of 13.4 kpc from the Sun, and the core radius is $1.19^\prime$ with the absolute visual magnitude being  -6.92 magnitudes (\citealt{Harris1996}, 2010 edition).

We see a bright radio source within the half-light radius of Pal 11. We detect this source in both the first and second epoch of VCSS. This source matches cataloged observations in the TGSSr, NVSS, RACS-low, and the first and second epochs of VLASS. The measurements are well-fit by a simple power law with a spectral index of $\alpha \sim -0.7$.

\subsection{NGC 2298}

NGC 2298 is at a distance of 10.8 kpc from the Sun, and the core radius is $0.31^\prime$ with the absolute visual magnitude being -6.31 magnitudes (\citealt{Harris1996}, 2010 edition).

We see a radio source within the half-light radius of NGC 2298. This source appears in the first and second epochs of VCSS as a single point source, and has a catalog match in NVSS, SUMSS, TGSSr, RACS-low, VLASS1 and VLASS2. In both epochs of VLASS, the source appears as a close double, and we summed the components to calculate a spectral index. We find a spectral index of $\alpha \sim 0.1$.

\subsection{\glimpse}

\glimpse\ was found by Spitzer during the Galactic Legacy Infrared Mid Plane Survey (\citealt{Kobulnicky2005}). Here we adopt a distance of 3.3 kpc from the Sun \citep{Hare2018}, and the core radius is $0.59^\prime$ with the absolute visual magnitude being -5.91 magnitudes (\citealt{Harris1996}, 2010 edition). 

We found a single source in archival VLITE A-configuration measurements from 2021 February 27. The data had a resolution of $4.7^{\prime\prime}\times5.0^{\prime\prime}$ at a position angle of $52^{o}$, and the source is detected at a peak significance of $6.4\sigma$.  No matches in published catalogs were found for this source; however it was clearly visible in the TGSSr.  Using PyBDSF to fit the position, we were able to measure the source flux density.  The spectral index between VLITE and the TGSSr was $\alpha=-2.8$.  Because of the extremely steep spectrum we obtained additional data on radio, X-ray and infrared wavelengths to further investigate the possibility that it was a pulsar.  Our full analysis is discussed in Section~\ref{sec:glimpse}, and all of our measurements for the source are tabulated in Table ~\ref{tab:GlimpseFlux}.

\subsection{Candidates at 3-5$\sigma$ with Detections at Other Frequencies}

\subsubsection{NGC 5053}
NGC 5053 is at a distance of 17.4 kpc from the Sun, and the core radius is $2.08^\prime$ with the absolute visual magnitude being -6.76 magnitudes (\citealt{Harris1996}, 2010 edition).

Cataloged only in VCSS1 at a marginal level of $3.8\sigma$, this source is not clearly seen in VCSS2.  It has matches in the NVSS, FIRST, and RACS-low catalogs. Although it is uncataloged, a source at this position is also visible in the TGSSr images. We used PyBDSF to fit the image at this position but found that the fitted size was adversely affected by the local noise. We thus forced the fitted size of the source in the TGSSr image to match the beam size.  The result of this fit is a source that has a peak SNR of 4.6 and a flux density of $17.1 \pm 6.3$ mJy.

The final spectrum is shown in Figure~\ref{fig:spectra}.  Although the TGSSr measurement suggests a slight turnover at low frequencies, the measurement has a large degree of uncertainty. So we have chosen to fit only a simple power-law, which has a spectral index of $\alpha = -1.3$.

\subsubsection{Pal 14}

Pal 14 is at a distance of 76.5 kpc from the Sun, and the core radius is $0.82^\prime$ with the absolute visual magnitude being -4.80 magnitudes (\citealt{Harris1996}, 2010 edition).

In VCSS1, we see a faint radio source within the half-light radius of the cluster. This source has a catalog match in NVSS, FIRST, RACS-low, and the second epoch of VLASS. Visually this source is seen in the VLASS1 image but appears to be uncataloged. It is cataloged in VLASS2 and included in spectrum. We find a spectral index of $\alpha \sim -1.4$

\subsubsection{NGC 6681}

NGC 6681 is at a distance of 9.0 kpc from the Sun, and the core radius is $0.03^\prime$ with the absolute visual magnitude being -7.12 magnitudes (\citealt{Harris1996}, 2010 edition).

In VCSS1 and VCSS2, a source appears within the half-light radius of the cluster. In VCSS2, the source appears as a close double, and has catalog matches in RACS and NVSS. It appears as a blended double in RACS-low, and is probably resolved out in VLASS.
An uncataloged SUMSS source is also present at this position. Fitting with PyBDSF gave a flux density of $8.4 \pm 1.9$ mJy at $7.6\sigma$. We find that this source has a spectral index of $\alpha \sim -0.4$.

\subsubsection{Arp 2}

Arp 2 is at a distance of 28.6 kpc from the Sun, and the core radius is $1.19^\prime$ with the absolute visual magnitude being -5.29 magnitudes (\citealt{Harris1996}, 2010 edition).

In VCSS1 and VCSS2, we detect a close double within the half-light radius. This source matches cataloged observations in NVSS, RACS, SUMSS, and the second epoch of VLASS. In RACS, the source also appears as a close a double, consistent with the VCSS morphology. A simple power law fit to the measurements gives a spectral index of  $\alpha \sim -1.4$.

\subsubsection{Terzan 3}
Terzan 3 is at a distance of 8.2 kpc from the Sun, and the core radius is $1.18^\prime$ with the absolute visual magnitude being -4.82 magnitudes (\citealt{Harris1996}, 2010 edition).

We detect a source in VCSS1 that appears to be a faint double. This source is also cataloged in TGSS, RACS, and NVSS. Visually inspecting VLASS images, hints of lobes are present but seem mostly resolved out. An uncataloged source is present in SUMSS at this position. It appears as a single component and fitting this source with PyBDSF gives a flux density of $14.9 \pm 4.0$ with a signal to noise ratio of 6.2. A simple power-law fit for this source has a spectral index of $\alpha \sim -0.8$

\begin{figure*}
    \centering
    \includegraphics[width=6.5in]{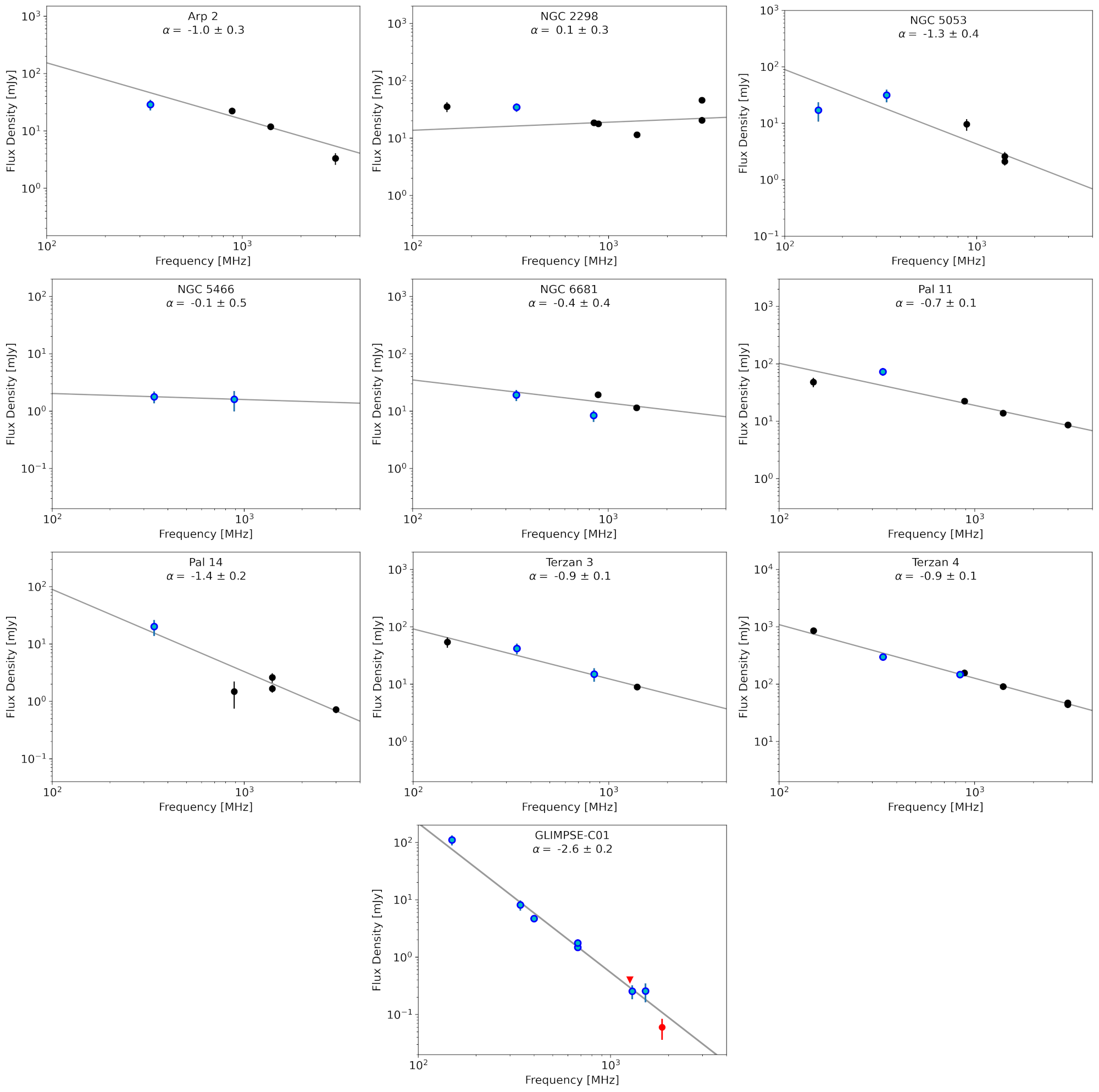}
    \caption{Spectra of VLITE-detected sources in globular clusters without known pulsars. Data presented in this work are indicated in blue; data from publicly available radio source catalogs are plotted in black. Red points were not used for fitting the source spectrum in GLIMPSE-C01.\label{fig:spectra}}
\end{figure*}

\section{Discussion}
\label{discussionm}

\subsection{Statistics}
\label{statistics}

We calculate the expected number of background sources we observe based on source counts from FIRST and RACS-low. Source fluxes were scaled to 340~MHz with spectral indices of $-0.71$ for FIRST and $-0.75$ for RACS-low. These indices were established as the median values among catalog sources matched to point sources of similar resolution in the VLITE data archive. We fit a power law within the scaled flux range $6-20$~mJy to estimate the source count function as $N(>S) = 194 S^{-0.72}$, where $N$ is the number of sources per square degree above flux density $S$ in mJy.

The expected count of background sources observed above a signal to noise ratio $X$ is obtained by summing the count for each cluster: $n = \Omega N(>X\sigma)$. $\sigma$ represents the image noise at the cluster's location. $\Omega$ is the solid angle subtended by the larger of the core or half-light radius. For the 97 clusters in our final sample, we searched a total solid angle of 0.13 square degrees. 

At our sensitivity limits, the expected number of background sources exceeding $5 \sigma$ is approximately four. Excluding the \glimpse\ pulsar, we detected four sources, in agreement with the expected outcome. At $>3 \sigma$, however, the predicted count of background sources is six, whereas we observed 10. The surplus of sources $3-5\sigma$ thus suggests that a portion is likely of Galactic origin and may include pulsar candidates. Additionally, we note that the discovery of bright background AGN behind globular clusters would potentially be useful for astrometry projects, and for probing the interstellar medium of clusters using the AGN as backlights \citep{vanLoon2009}.

\subsection{Discussion of Pulsar in \glimpse}
\label{sec:glimpse}
\glimpse\ is located at a Galactic latitude of only $-0.1^{\circ}$, and optical and infrared observations have found that it has differential reddening across it \citep{Hare2018}. Despite its location in the Galactic Plane, it shares many properties with the older globular clusters typically seen in the Galactic Halo.  Its age and distance are both uncertain, and its stellar main sequence is consistent with either a young or intermediate aged cluster \citep{Hare2018}. Similarly, its distance from us is uncertain and estimates between $3$ and $5$ kpc are found in the literature; in common with other recent studies we adopt a distance of $3.3$ kpc \citep{Hare2018}.  

The pulsar candidate discovered in the VLITE image is the same source recently reported as a flat spectrum blazar candidate in \citet{Luque-Escamilla2023} based on measurements in archival radio data from the TGSSr (150 MHz) and the MAGPIS (90cm/350 MHz; \citealt{Helfand2006}) radio surveys. They find that it is coincident with a variable X-ray source in the {\it Chandra} data. The source is not seen in the higher frequency (20cm) images of MAGPIS, with a limit that requires a spectral index of $\alpha<-1.1$. \citet{Luque-Escamilla2023} suggest that this is due to temporal variability of the radio flux density.  However, we note that the TGSSr data at this position were recorded in 2010, while the MAGPIS 90cm data at this position were observed in 2001. If the source is variable, then the apparent ``flat" spectral index would not be a valid measure of the intrinsic source spectrum.  

The MAGPIS 90cm image does show a $\sim$100 mJy/bm source roughly coincident with the pulsar.  However the beam is roughly $1^{\prime}$, which covers nearly the entire cluster, and it is clear the ``source" is embedded in a larger diffuse emission structure that extends well beyond the cluster itself.  As such, it is difficult to compare directly to the higher resolution TGSSr ($25^{\prime\prime}$) and VLITE data.  The VLITE image is from nearly two hour of data taken on 27 February 2021, and has a resolution of $6^{\prime\prime}$.  There is no evidence of diffuse emission at the position of the cluster.  The flux density of the source ($S_{VLITE} \sim 8.1$ mJy) is lower than the reported 350 MHz MAGPIS flux by a factor of more than 10. While it is possible that the source is variable by such a large amount, it seems more likely that the MAGPIS image is confused by the visible Galactic structures.

Based on the tentative blazar association, this source was suggested by \citet{Luque-Escamilla2023} to be the counterpart of a Fermi source with an overlapping source error circle.  In the context of the pulsar interpretation, this association is extremely unlikely, given that the inferred spin-down luminosity from the X-ray data is smaller than the gammma-ray luminosity if the gamma-ray source is in GLIMPSE-C01.

In Figure~\ref{fig:spectra} we have plotted all of our measurements for the flux density of this pulsar, spanning frequencies from 150 MHz (TGSSr) to 2 GHz (GBT pulse flux).  Tabulated information on these measurements is presented in Table~\ref{tab:GlimpseFlux}.  We note that the measurements span more than a decade of observations, but the spectrum is well fit by an unbroken power law with a spectral index of $\alpha \sim -2.6$.  

Hard (2-10 keV) X-ray luminosities, $L_X$ for millisecond pulsars are typically about 0.1\% of the spin-down luminosities, $\dot{E}$ \citep{Seward,Becker}, with some evidence for a steeper-than-linear overall dependence of X-ray luminosity on spin-down luminosity\citep{Possenti2002}.  For sources like \glimpse, with large absorption columns, the softer thermal emission from the polar caps \citep{Bogdanov} is negligible.  For the estimated distance of 3.3~kpc, the flux estimate corresponds to a luminosity of $6\times10^{30}$ erg/sec.  

Interestingly, the pulsar in \glimpse\ has a higher $L_X$ than most millisecond pulsars in globular clusters, while also having a slower spin period, suggesting a high magnetic field.  The expectation of a high magnetic field is independent of specific prescriptions for the relation between $\dot{E}$ and $L_X$; under the assumption of the linear relation between $\dot{E}$ and $L_X$ \citep{Becker}, the magnetic field $B$ can be expected to scale as $\dot{E}^{1/2}P^2$.  For a neutron star to have a spin down luminosity of $6\times10^{33}$ for a period of 19.78 msec requires a period derivative of about $10^{-18}$ s/s \citep{2016era..book.....C}, and a magnetic field of about $4.4\times10^9$ G \citep{2016era..book.....C}.  The characteristic age of the pulsar is then about $3\times10^8$ yr, suggesting that it was spun up relatively recently.  

A caveat here is that many of the X-ray bright millisecond pulsars in globular clusters are ``spider" binaries, with the X-rays coming from a shock between the pulsar wind and the outflow of the companion star \citep{ZhaoHeinke}.  These systems are generally very nearly Roche lobe-filling, which is unlikely for our object given that in over 7 hours of pulsar data, it shows evidence for acceleration in a binary, but not for the change in acceleration that would be expected for an orbit of less than a day.  While it is possible for redbacks to reach periods at least as long as 2 days \citep{PichardoMarcano2021}, the combination of a spin period longer than 10 msec and a period longer than a day would be outside the known range for redback systems, while having a relatively high magnetic field for a partially recycled pulsar would not be surprising.  We thus presume that the $B\sim10^9$ G, $\tau\sim10^8$ yr scenario is much more likely.

\begin{figure}
\plotone{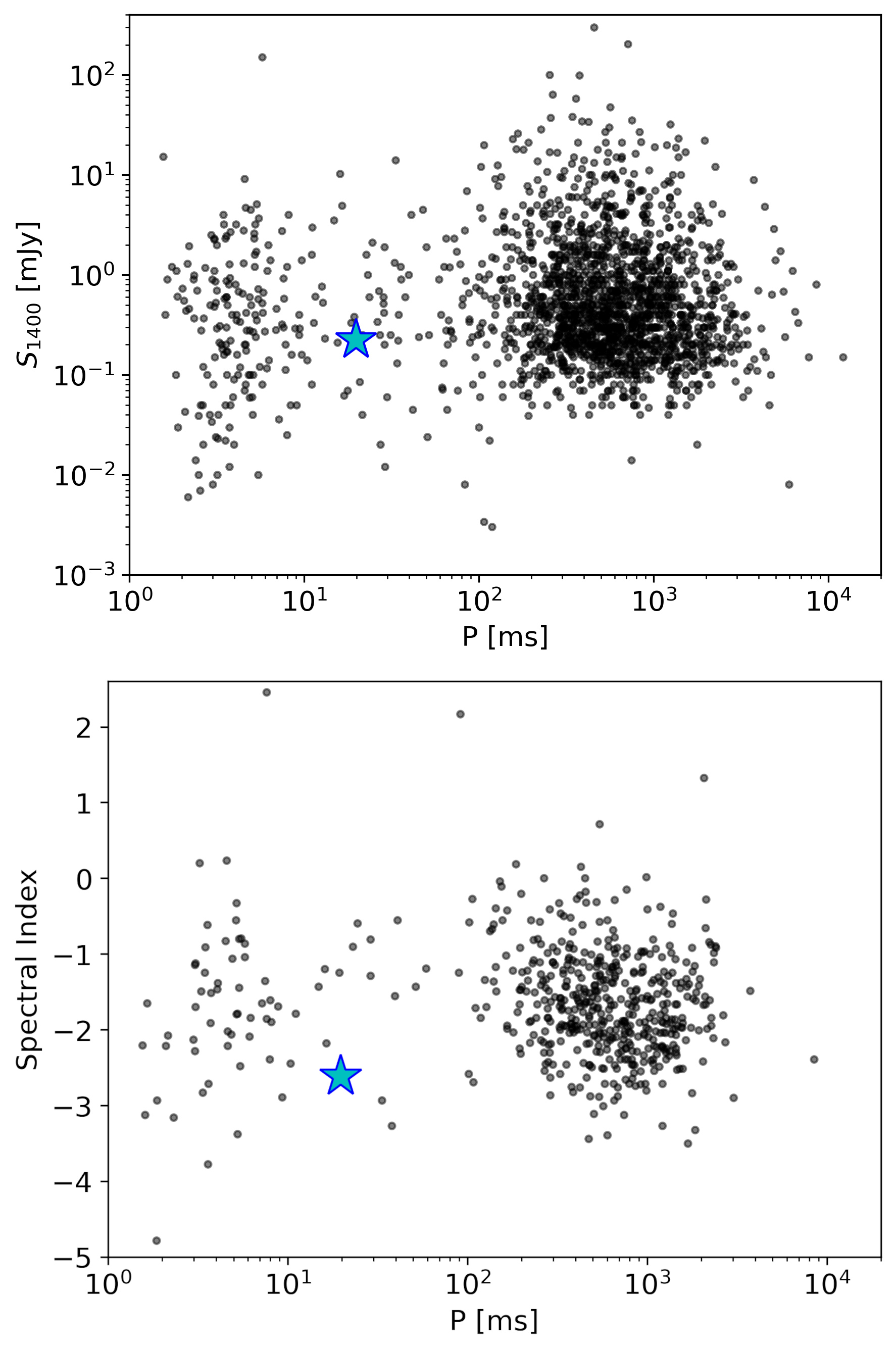}
\caption{Spin period and 1.4 GHz flux density (top) and spectral index (bottom) of pulsars in the ATNF Catalog. The newly discovered MSP in \glimpse\ is indicated by the star symbol.\label{fig:psrs}}
\end{figure}

The most likely implication of this intermediate magnetic field, along with a spin period that is slow relative to millisecond pulsars, but faster than the Crab, is that the spin-up of this pulsar was interrupted, as illustrated in Figure~\ref{fig:psrs}.  A class of pulsars with young characteristic ages and a wide range of spin period exists in Galactic globular clusters and is explained by disruption of the X-ray binary during the accretion phase \citep{VerbuntFreire}: the magnetic field of the newly formed radio pulsar reflects how advanced was the degradation of the magnetic field that is thought to happen during the recycling process.  Given the presence of a binary companion for the pulsar in \glimpse, we can expect that the progenitor X-ray binary must have been disrupted by an exchange encounter, rather than an encounter that ionized the binary.  
These partially recycled pulsars are found predominantly in clusters with relatively high rates of stellar interactions per star \citep{VerbuntFreire};  whether this is also the case for \glimpse\, is still unclear because of its high reddening, uncertain distance, uncertain age, and heavy contamination from background stars \citep{Davies2011,
Hare2018, Bahramian2023}.  Future JWST data with proper motion cleaning could help solve this problem.

An alternative possibility cannot be excluded at the present time.  The X-ray source could potentially be a chance counterpart to the radio continuum source, or the radio continuum source could potentially be associated with a faster pulsar that has not yet been detected. Given the source's location on the outskirts of the cluster core, where there are few similarly bright sources, the chance superposition is unlikely, and the intermediate spin periods should typically lead to higher values of $L_X$ than are usually found for globular cluster pulsars.  The similar flux density for the pulsed emission and the emission from imaging make it unlikely that the ``wrong" pulsar has been found.  Still, until the pulsar can be well-localized, a definitive association with the X-ray source cannot be made.

\section{Conclusions and Future Work}

 Our radio imaging search for pulsars in a sample of 97 globular clusters with no known pulsars has identified 10 sources within the larger of the core or half-light search radius for these systems. Based on extragalactic source counts, we have an excess of four sources over predictions for our search area given our sensitivity limits, indicating that we are likely sensitive to sources of Galactic origin. For one of our pulsar candidates, in \glimpse, a detailed analysis of archival and new observations has allowed us to confirm the first millisecond pulsar (PSR\ J1848-0129A) in this globular cluster. We determine the radio source spectral index for PSR\ J1848-0129A as $\alpha = -2.6\pm 0.2$ and measure the spin period as 19.78 ms.
 
 In the short term, it will be important to carry out regular timing and establish an orbital and timing solution for \glimpse A. A timing solution will provide a very precise position, which will be important to confirm the association. The timing will also measure the spin-down rate, which will eventually provide the characteristic age of the system. Determining the orbital parameters of the system might confirm a recent secondary exchange interaction, which would be of great interest.

This work provides a new example of the efficacy of spectral index searches for pulsars.  In the SKA era, deep, wide-field surveys of the Southern sky will become commonplace.  This will prove useful both for finding more globular cluster pulsars and for finding more pulsars deep in the Galactic Plane, the key set of pulsars for doing things like mapping out the Milky Way's magnetic field \citep{2008MNRAS.386.1881N}.  

Deeper observations of globular clusters, both with VLITE and high frequency VLA data, or with dedicated P-band observations (both with the current VLA P-band system and the GMRT, and with future arrays like SKA) could be helpful, as well.  \citet{2004MNRAS.348.1409M} find that the faintest pulsars in 47~Tuc are likely to have pseudoluminosities of about 0.4 mJy kpc$^2$ at 1.4 GHz, which corresponds to about 3 mJy kpc$^2$ at 350 MHz for a spectral index of $-1.4$.  At a distance of 10 kpc, this corresponds to a flux density of about 30 $\mu$Jy.   This is a factor of about 100-500 deeper than the typical observation presented in this paper, which means that continuing to obtain deep A-config P-band data is likely to produce many more pulsar discoveries via imaging.  At the same time, it will take SKA, with long baselines, to detect all the Milky Way's pulsars via imaging.

\begin{acknowledgements}
AM received support through the Naval Research Enterprise Internship Program (NREIP) to undertake this research at the U.S.\ Naval Research Laboratory. Basic research in radio astronomy at the U.S.\ Naval Research Laboratory is supported by 6.1 Base funding. Construction and installation of VLITE was supported by the NRL Sustainment Restoration and Maintenance fund. The National Radio Astronomy Observatory is a facility of the National Science Foundation operated under cooperative agreement by Associated Universities,  Inc. We thank the staff of the GMRT that made the observations possible. GMRT is run by the National Centre for Astrophysics of the Tata Institute of Fundamental Research. This research has made use of ``Aladin sky atlas" developed at CDS,  Strasbourg Observatory,  France.  We thank F. Schinzel (NRAO) for assistance with archival JVLA data during the early parts of this study.   We thank Dale Frail and Miller Goss for valuable discussions of the history of spectral-index-based searches for pulsars, and TJM thanks the MAVERICS team for useful discussions.

\end{acknowledgements}

\vspace{5mm}
\facilities{VLA (NRAO), GBT, GMRT, {\it Chandra}}

\bibliography{PSRGlimpse}{}
\bibliographystyle{aasjournal}

\begin{longtable*}{lcccccccc}
\caption{Globular Cluster sample searched.}\label{tab:sample}\\
\hline
Cluster & RA & Dec & Half-light Radius$^{a}$ & Core Radius$^{a}$  & Best Image & Resolution & $3\sigma$\\
Name & (Deg, J2000) & (Deg, J2000) & (arcmin) & (arcmin) & Source & (arcsec) & (mJy bm$^{-1}$)  \\
\hline
\endhead
\hline
\endfoot
1636-283 & 249.85604 & -28.39869 & 0.5 & 0.5 & VCSS1 & 15.6 & 11.1 \\ 
2MS-GC01 & 272.09087 & -19.82972 & 1.65 & 0.85 & VLITE & 7.2 & 3.9 \\ 
2MS-GC02 & 272.40208 & -20.77889 & 0.55 & 0.55 & VLITE & 7.2 & 6.0 \\ 
AM 4 & 209.09042 & -27.1675 & 0.43 & 0.41 & VLITE & 16.2 & 5.2 \\ 
Arp 2 & 292.18379 & -30.35564 & 1.77 & 1.19 & VCSS2 & 20.0 & 16.7 \\ 
BH 261 & 273.5275 & -28.635 & 0.55 & 0.4 & VCSS2 & 15.6 & 14.5 \\ 
Djorg 1 & 266.86792 & -33.06556 & 1.59 & 0.5 & VCSS1 & 20.0 & 17.0 \\ 
Djorg 2 & 270.45458 & -27.82583 & 1.05 & 0.33 & VLITE & 6.6 & 3.1 \\ 
Eridanus & 66.18542 & -21.18694 & 0.46 & 0.25 & VCSS2 & 20.0 & 13.7 \\ 
GLIMPSE-C01 & 282.20708 & -1.49722 & 0.65 & 0.59 & VLITE & 5.0 & 3.6 \\ 
GLIMPSE-C02 & 274.62708 & -16.97722 & -- & 0.7 & VCSS1 & 20.0 & 27.4 \\ 
HP 1 & 262.77167 & -29.98167 & 3.1 & 0.03 & VLITE & 8.8 & 8.6 \\ 
IC 1257 & 261.78542 & -7.09306 & 1.4 & 0.25 & VCSS1 & 21.9 & 14.8 \\ 
IC 1276 & 272.68417 & -7.20761 & 2.38 & 1.01 & VCSS1 & 26.2 & 16.2 \\ 
Ko 1 & 179.82708 & 12.26 & 0.26 & 0.33 & VLITE & 6.1 & 7.4 \\ 
Ko 2 & 119.57083 & 26.255 & 0.21 & 0.25 & VLITE & 20.3 & 7.0 \\ 
Liller 1 & 263.35208 & -33.389 & -- & 0.06 & VLITE & 14.7 & 8.2 \\ 
NGC 1904 & 81.04621 & -24.52472 & 0.65 & 0.16 & VCSS2 & 20.0 & 15.3 \\ 
NGC 2298 & 102.24754 & -36.00531 & 0.98 & 0.31 & VCSS1 & 20.4 & 12.3 \\ 
NGC 2419 & 114.53529 & 38.88244 & 0.89 & 0.32 & VLITE & 17.7 & 7.9 \\ 
NGC 288 & 13.1885 & -26.58261 & 2.23 & 1.35 & VLITE & 9.2 & 1.2 \\ 
NGC 4147 & 182.52625 & 18.54264 & 0.48 & 0.09 & VLITE & 5.6 & 5.6 \\ 
NGC 4590 & 189.86658 & -26.74406 & 1.51 & 0.58 & VCSS1 & 16.0 & 13.5 \\ 
NGC 5053 & 199.11288 & 17.70025 & 2.61 & 2.08 & VLITE & 17.0 & 7.8 \\ 
NGC 5466 & 211.36371 & 28.53444 & 2.3 & 1.43 & VLITE & 6.8 & 0.7 \\ 
NGC 5634 & 217.40512 & -5.97642 & 0.86 & 0.09 & VCSS1 & 24.8 & 10.9 \\ 
NGC 5694 & 219.90121 & -26.53894 & 0.4 & 0.06 & VCSS1 & 17.2 & 15.6 \\ 
NGC 5824 & 225.99429 & -33.06822 & 0.45 & 0.06 & VCSS1 & 19.7 & 17.6 \\ 
NGC 5897 & 229.35208 & -21.01028 & 2.06 & 1.4 & VLITE & 7.2 & 7.8 \\ 
NGC 6093 & 244.26004 & -22.97608 & 0.61 & 0.15 & VCSS1 & 15.5 & 20.1 \\ 
NGC 6139 & 246.91821 & -38.84875 & 0.85 & 0.15 & VCSS1 & 24.2 & 13.6 \\ 
NGC 6144 & 246.80775 & -26.0235 & 1.63 & 0.94 & VLITE & 6.5 & 2.0 \\ 
NGC 6171 & 248.13275 & -13.05378 & 1.73 & 0.56 & VCSS1 & 19.4 & 34.8 \\ 
NGC 6229 & 251.74496 & 47.52775 & 0.36 & 0.12 & VLITE & 4.3 & 3.5 \\ 
NGC 6235 & 253.35546 & -22.17744 & 1.0 & 0.33 & VCSS1 & 21.0 & 22.8 \\ 
NGC 6256 & 254.88592 & -37.12139 & 0.86 & 0.02 & VCSS1 & 24.2 & 17.2 \\ 
NGC 6273 & 255.6575 & -26.26797 & 1.32 & 0.43 & VLITE & 9.1 & 3.5 \\ 
NGC 6284 & 256.11879 & -24.76486 & 0.66 & 0.07 & VCSS1 & 19.9 & 26.2 \\ 
NGC 6287 & 256.28804 & -22.70836 & 0.74 & 0.29 & VCSS1 & 21.0 & 38.4 \\ 
NGC 6293 & 257.5425 & -26.58208 & 0.89 & 0.05 & VLITE & 8.4 & 2.9 \\ 
NGC 6304 & 258.63437 & -29.46203 & 1.42 & 0.21 & VCSS2 & 20.0 & 15.9 \\ 
NGC 6316 & 259.15542 & -28.14011 & 0.65 & 0.17 & VCSS1 & 19.9 & 17.0 \\ 
NGC 6325 & 259.49671 & -23.766 & 0.63 & 0.03 & VCSS1 & 21.0 & 20.7 \\ 
NGC 6333 & 259.79692 & -18.51594 & 0.96 & 0.45 & VCSS1 & 16.8 & 24.6 \\ 
NGC 6355 & 260.99412 & -26.35342 & 0.88 & 0.05 & VCSS1 & 19.9 & 16.1 \\ 
NGC 6356 & 260.89554 & -17.81303 & 0.81 & 0.24 & VLITE  &  6.8 & 24.1 \\
NGC 6366 & 261.93433 & -5.07986 & 2.92 & 2.17 & VCSS1 & 21.3 & 16.2 \\ 
NGC 6380 & 263.61667 & -39.06917 & 0.74 & 0.34 & VCSS1 & 23.5 & 16.2 \\ 
NGC 6401 & 264.6525 & -23.9095 & 1.91 & 0.25 & VCSS1 & 16.9 & 20.3 \\ 
NGC 6426 & 266.22771 & 3.17014 & 0.92 & 0.26 & VCSS1 & 19.3 & 14.6 \\ 
NGC 6453 & 267.71542 & -34.59917 & 0.44 & 0.05 & VCSS1 & 17.4 & 13.9 \\ 
NGC 6528 & 271.20683 & -30.05628 & 0.38 & 0.13 & VLITE & 10.4 & 4.9 \\ 
NGC 6535 & 270.96046 & -0.29764 & 0.85 & 0.36 & VCSS1 & 19.2 & 16.3 \\ 
NGC 6540 & 271.53583 & -27.76528 & -- & 0.03 & VLITE & 10.8 & 5.3 \\ 
NGC 6553 & 272.32333 & -25.90869 & 1.03 & 0.53 & VLITE & 10.7 & 3.0 \\ 
NGC 6558 & 272.57333 & -31.76389 & 2.15 & 0.03 & VCSS2 & 20.0 & 15.3 \\ 
NGC 6569 & 273.41167 & -31.82689 & 0.8 & 0.35 & VCSS2 & 20.0 & 14.8 \\ 
NGC 6637 & 277.84625 & -32.34808 & 0.84 & 0.33 & VCSS2 & 20.0 & 16.7 \\ 
NGC 6638 & 277.73375 & -25.49747 & 0.51 & 0.22 & VCSS1 & 20.0 & 17.0 \\ 
NGC 6642 & 277.97542 & -23.47519 & 0.73 & 0.1 & VLITE & 9.0 & 4.9 \\ 
NGC 6681 & 280.80317 & -32.29211 & 0.71 & 0.03 & VCSS1 & 20.0 & 20.0 \\ 
NGC 6715 & 283.76387 & -30.47986 & 0.82 & 0.09 & VCSS2 & 20.0 & 13.3 \\ 
NGC 6717 & 283.77517 & -22.70147 & 0.68 & 0.08 & VCSS1 & 20.9 & 21.5 \\ 
NGC 6723 & 284.88813 & -36.63225 & 1.53 & 0.83 & VCSS2 & 20.0 & 18.5 \\ 
NGC 6779 & 289.14821 & 30.18347 & 1.1 & 0.44 & VCSS2 & 20.0 & 34.5 \\ 
NGC 6809 & 294.99879 & -30.96475 & 2.83 & 1.8 & VCSS2 & 20.0 & 13.4 \\ 
NGC 6864 & 301.51954 & -21.92117 & 0.46 & 0.09 & VCSS2 & 20.0 & 15.4 \\ 
NGC 6934 & 308.54737 & 7.40447 & 0.69 & 0.22 & VCSS1 & 17.6 & 10.1 \\ 
NGC 6981 & 313.36542 & -12.53731 & 0.93 & 0.46 & VCSS1 & 24.8 & 9.6 \\ 
NGC 7006 & 315.37242 & 16.18733 & 0.44 & 0.17 & VCSS1 & 18.8 & 10.9 \\ 
NGC 7492 & 347.11096 & -15.6115 & 1.15 & 0.86 & VCSS1 & 21.9 & 13.3 \\ 
Pal 1 & 53.3335 & 79.58106 & 0.46 & 0.01 & VCSS1 & 23.9 & 12.4 \\ 
Pal 10 & 289.50875 & 18.57167 & 0.99 & 0.81 & VLITE & 17.4 & 8.0 \\ 
Pal 11 & 296.31 & -8.00722 & 1.46 & 1.19 & VCSS1 & 21.8 & 15.8 \\ 
Pal 12 & 326.66183 & -21.25261 & 1.72 & 0.02 & VCSS1 & 20.7 & 9.4 \\ 
Pal 13 & 346.68517 & 12.772 & 0.36 & 0.48 & VLITE & 5.1 & 2.1 \\ 
Pal 14 & 242.7525 & 14.95778 & 1.22 & 0.82 & VCSS1 & 14.9 & 13.3 \\ 
Pal 15 & 254.9625 & -0.53889 & 1.1 & 1.2 & VCSS1 & 20.5 & 17.1 \\ 
Pal 2 & 71.52463 & 31.3815 & 0.5 & 0.17 & VCSS2 & 20.7 & 29.0 \\ 
Pal 3 & 151.38292 & 0.07167 & 0.65 & 0.41 & VCSS1 & 20.5 & 18.5 \\ 
Pal 4 & 172.32 & 28.97358 & 0.51 & 0.33 & VCSS1 & 14.1 & 13.8 \\ 
Pal 5 & 229.02187 & -0.11161 & 2.73 & 2.29 & VLITE & 18.6 & 14.7 \\ 
Pal 6 & 265.92583 & -26.2225 & 1.2 & 0.66 & VLITE & 5.8 & 7.7 \\ 
Pal 8 & 280.37458 & -19.82583 & 0.58 & 0.56 & VCSS1 & 17.2 & 27.8 \\ 
Pyxis$^{b}$ & 136.99083 & -37.22139 & -- & -- & VCSS2 & 20.0 & 16.0 \\ 
Terzan 10 & 270.90167 & -26.0725 & 1.55 & 0.9 & VCSS1 & 15.4 & 24.2 \\ 
Terzan 12 & 273.06583 & -22.74194 & 0.75 & 0.83 & VCSS1 & 15.6 & 39.8 \\ 
Terzan 2 & 261.88792 & -30.80233 & 1.52 & 0.03 & VCSS1 & 20.0 & 19.6 \\ 
Terzan 3 & 247.167 & -35.35347 & 1.25 & 1.18 & VCSS1 & 27.1 & 14.0 \\ 
Terzan 4 & 262.6625 & -31.59553 & 1.85 & 0.9 & VCSS1 & 20.0 & 27.7 \\ 
Terzan 6 & 267.69325 & -31.27539 & 0.44 & 0.05 & VLITE & 6.9 & 2.6 \\ 
Terzan 7 & 289.433 & -34.65772 & 0.77 & 0.49 & VCSS1 & 16.7 & 13.8 \\ 
Terzan 8 & 295.43504 & -33.99947 & 0.95 & 1.0 & VCSS1 & 17.0 & 11.5 \\ 
Terzan 9 & 270.41167 & -26.83972 & 0.78 & 0.03 & VLITE & 6.6 & 5.3 \\ 
Ton 2 & 264.04375 & -38.55333 & 1.3 & 0.54 & VLITE & 15.9 & 5.8 \\ 
UKS 1 & 268.61333 & -24.14528 & -- & 0.15 & VCSS1 & 16.9 & 23.4 \\ 
Whiting 1 & 30.7375 & -3.25278 & 0.22 & 0.25 & VLITE & 5.5 & 3.4 \\
\hline

\multicolumn{8}{l}{$^{a}$ We use a -- to indicate where no half-light or core radius is known.}\\
\multicolumn{8}{l}{$^{b}$ No known radii were available. A search radius of $3^\prime$ was used, chosen based on the largest radius observed within our sample.}\\
\hline

\end{longtable*}

\pagebreak

\begin{deluxetable*}{lcccccccc}
\label{tab:Sources1}
\tablecaption{VLITE detections within globular clusters}
\tablehead{
\colhead{Cluster} & \colhead{Source Name} & \colhead{RA} & \colhead{Dec} & \colhead{Peak Flux} & \colhead{Total Flux} & \colhead{SNR} & \colhead{Catalog} & \colhead{Spectral Index}\\
\colhead{Name} & \colhead{} & \colhead{(Deg, J2000)} & \colhead{(Deg, J2000)} & \colhead{(mJy bm$^{-1}$)} & \colhead{(mJy)} & \colhead{} & \colhead{Matches} & \colhead{}}
\startdata
\multicolumn{9}{c}{VLITE detections with SNR $> 5$}\\
\hline
Terzan 4 & VCSS1 J173033.4-313456 & 262.63921 & -31.58215 & $242.5 \pm 5.6$ & $247.0 \pm 9.8$ & 43.5 & 1, 2$^*$, 3, 4, 6, 7 & $-0.9 \pm 0.1$ \\
NGC 5466 & VLITE-A J140527.6+283100 & 211.36502 & 28.51671 & $2.1 \pm 0.2$ & $1.8 \pm 0.3$ & 9.6 & 3$^*$ & $-0.1 \pm $ 0.5\\
Pal 11 & VCSS1 J194508.8-080010 & 296.28651 & -8.00264 & $27.8 \pm 3.6$ & $46.8 \pm 9.2$ & 7.7 & 1, 3, 4, 6, 7 & $-0.7 \pm $ 0.1\\
NGC 2298 & VCSS1 J064857.9-360035 & 102.24120 & -36.00986 & $19.4 \pm 2.6$ & $22.0 \pm 5.1$ & 7.4 & 1, 2, 3, 4, 6, 7 & $-0.1 \pm 0.3$\\
\glimpse\ & VLITE-A J184848.1-012958 & 282.20071 & -1.49970 & $7.7 \pm 1.2$ & $7.3 \pm 2.0$ & 6.4 & 1$^*$ & $-2.6 \pm 0.2$ \\
\hline
\multicolumn{9}{c}{Marginal VLITE detections with matches in other catalogs}\\
\hline
NGC 5053 & VLITE-A J131619.8+174340 &  199.08238 & 17.72770 & $9.2 \pm 2.6$ & $19.1 \pm 7.6$ & 3.8 & 1$^*$, 3, 4, 5 & $-1.3 \pm 0.4$\\
Pal 14 & VCSS1 J161057.8+145706 &  242.74078 & 14.95160 & $8.8 \pm 2.7$ & $12.1 \pm 6.0$ & 3.3 & 3, 4, 5, 7 & $-1.4 \pm 0.2$\\
NGC 6681 & VLITE-A J184311.9-321654 &  280.79941 & -32.28157 & $17.3 \pm 3.0$ & $10.5 \pm 3.9$ &  4.8 & 2$^*$, 3, 4 & $-0.4 \pm 0.4$\\
Arp 2 & VCSS2 J192847.2-302239 & 292.19661 & -30.37739 & $15.2 \pm 3.2$ & $15.9 \pm 5.7$ & 4.8 & 2, 3, 4, 7 & $-1.4 \pm 0.1$\\
Terzan 3 & VCSS1 J162835.2-352119 & 247.14665 & -35.35534 & $14.3 \pm 3.0$ & $25.8 \pm 8.2$ & 4.8 & 1, 2$^*$, 3, 4 & $-0.8 \pm 0.2$\\
\enddata
\tablecomments{Catalog matches refer to: 1 TGSSr \citep{tgss}, 2 SUMSS \citep{Mauch2003}, 3 RACS-low \citep{Hale_2021}, 4 NVSS \citep{Condon1998}, 5 FIRST \citep{1995ApJ...450..559B}, 6 VLASS epoch 1 \citep{lacy+20}, 7 VLASS epoch2 \citep{lacy+20}. Catalogs with $^*$ indicate source was visible but not included in the published catalog. Spectral index and errors were determined from fits shown in Figure~\ref{fig:spectra}.}
\end{deluxetable*}

\begin{deluxetable*}{lcccc}
\label{tab:GlimpseFlux}
\tablecaption{Fitted flux densities for the pulsar candidate in \glimpse.  The reported flux density for the GBT is the pulse flux; all others are continuum. 
}
\tablehead{\colhead{Dataset} & \colhead{Frequency} & \colhead{Flux density} & \colhead{Date} & \colhead{Reference}\\
\colhead{} & \colhead{(MHz)} & \colhead{(mJy)} & \colhead{} & \colhead{}
}
\startdata
TGSS-ADR1 & 150 MHz &  $71 \pm 9$  &  May 2010 & \citet{tgss} \\
VLITE & 340 MHz &  $8.1 \pm 1.1$  &  Feb 2021 & this work \\
GMRT & 400 MHz &  $3.8 \pm 0.2$  & Jan 2023 & this work \\
~ & 675 MHz & $1.3 \pm 0.1$ & Jul 2022 & this work \\
~ & 675 MHz & $1.1 \pm 0.1$ & Jan 2023 & this work \\
~ & 1280 MHz & $<0.4$ & Jan 2023 & this work \\ 
JVLA & 1296 MHz & $0.253 \pm 0.070$ & Aug 2023 & this work \\
~ & 1520 MHz & $0.269 \pm 0.055$ & May 2023 & this work \\
~ & 1840 MHz & $0.089 \pm 0.051$ & Aug 2023 & this work \\
GBT & 1850 MHz &  $0.06 \pm 0.02$ & Aug 2005 & this work \\
\enddata
\end{deluxetable*}

\begin{deluxetable*}{lccccccccccc}
\label{tab:counterparts}
\tablecaption{NIR counterparts to X-ray sources in \glimpse\ and their NIR magnitudes\label{mags} in the WFC3 IR and UVIS filters.}
\tablehead{
\colhead{Source} & \colhead{RA} & \colhead{DEC} & \colhead{F814W} &  \colhead{$\sigma_{F814W}$} & \colhead{F127M} &  \colhead{$\sigma_{127M}$} & \colhead{F139M} &  \colhead{$\sigma_{F139M}$} & \colhead{F153M} &  \colhead{$\sigma_{F153M}$} & \colhead{offset} 
}
\startdata
C1 & 282.201019 & -1.499609 & 18.153 & 0.003 & 16.460 & 0.002 & 16.206 & 0.002 & 15.995 & 0.002 & 0.31 \\
C2 & 282.200913& -1.499543 & ... & ... & 19.956 & 0.015 & 19.168 & 0.011 & 18.499 & 0.009 & 0.30 \\
\enddata
\tablecomments{All magnitudes are in VEGAMAG system.}
\end{deluxetable*}

\end{document}